\documentclass[graybox]{svmult}

\usepackage{mathptmx}       
\usepackage{helvet}         
\usepackage{courier}        
\usepackage{type1cm}        
\usepackage{makeidx}         
\usepackage{graphicx}        
\usepackage{multicol}        
\usepackage[bottom]{footmisc}

\makeindex             

\begin{document}


\newcommand{\chandra}{{\em Chandra}}
\newcommand{\xmmn}{{\em XMM-Newton}}
\newcommand{\spitzer}{{\em Spitzer}}
\newcommand{\xmm}{{\em XMM}}
\newcommand{\fermi}{{\em Fermi}}
\newcommand{\rosat}{{\em ROSAT}}

\newcommand{\apj}{ApJ}
\newcommand{\aj}{AJ}
\newcommand{\apjl}{ApJL}
\newcommand{\mnras}{MNRAS}
\newcommand{\aap}{A\&A}

\def\pwn1640{{HESS~J1640$-$465}}
\def\snr338{{G338.3$-$0.0}}
\def\velax{{Vela~X}}
\def\kms{km~s$^{-1}$}
\def\etal{{\rm et~al.\ }}


\title*{Multiwavelength Observations of Pulsar Wind Nebulae}
\author{Patrick Slane}
\institute{Patrick Slane \at Harvard-Smithsonian Center for Astrophysics, 
60 Garden St., Cambridge, MA 02138 (USA) \at \email{slane@cfa.harvard.edu}}
%
%
\maketitle

\abstract*{
The extended nebulae formed as pulsar winds expand into their
surroundings provide information about the composition of the winds,
the injection history from the host pulsar, and the material into
which the nebulae are expanding. Observations from across the
electromagnetic spectrum provide constraints on the evolution of
the nebulae, the density and composition of the surrounding ejecta,
the geometry of the central engines, and the long-term fate of the
energetic particles produced in these systems. Such observations
reveal the presence of jets and wind termination shocks, time-varying
compact emission structures, shocked supernova ejecta, and newly
formed dust. Here I provide a broad overview of the structure of
pulsar wind nebulae, with specific examples from observations
extending from the radio band to very-high-energy $\gamma$-rays that
demonstrate our ability to constrain the history and ultimate fate
of the energy released in the spin-down of young pulsars.
}
\abstract{
The extended nebulae formed as pulsar winds expand into their
surroundings provide information about the composition of the winds,
the injection history from the host pulsar, and the material into
which the nebulae are expanding. Observations from across the
electromagnetic spectrum provide constraints on the evolution of
the nebulae, the density and composition of the surrounding ejecta,
the geometry of the central engines, and the long-term fate of the
energetic particles produced in these systems. Such observations
reveal the presence of jets and wind termination shocks, time-varying
compact emission structures, shocked supernova ejecta, and newly
formed dust. Here I provide a broad overview of the structure of
pulsar wind nebulae, with specific examples from observations
extending from the radio band to very-high-energy $\gamma$-rays that
demonstrate our ability to constrain the history and ultimate fate
of the energy released in the spin-down of young pulsars.
}

\section{Introduction}
\label{sec:1}
The basic structure of a pulsar wind nebula is determined by the
spin-down energy injected by the central pulsar and the interaction
of the nebula with the interior regions of the supernova remnant
(SNR) in which it evolves. Losses from synchrotron radiation in the
nebular magnetic field, whose strength depends both on the nature
of the injected wind and on the evolving size of the PWN, inverse-Compton
(IC) scattering of ambient photons by the energetic electron
population within the nebula, and adiabatic expansion as the nebula
sweeps up the surrounding supernova ejecta, all combine to determine
the emission structure and long-term evolution of the nebula. (See
\cite{gae06} for a review.) Multiwavelength observations
of PWNe provide crucial information on the underlying particle
spectrum and strongly constrain both the magnetic field strength
and the stage of evolution. Of particular interest is the spectrum
of low-energy particles contained in the PWN. These retain the
history of early energy losses as well as possible signatures of
features in the pulsar injection spectrum.

Complex structure in PWN spectra can originate in a number of ways
that are associated with the long-term evolution as well
\cite{ren84,gsz09}.  In particular, when the reverse shock
compresses the PWN, the resulting increase in the magnetic field
initiates a new epoch of fast synchrotron cooling that adds to the
population of low energy particles. Models which do not account for
the reverse shock interaction will thus underestimate this population
and its associated radiation.  In addition, recent studies of the
spectrum immediately downstream of the wind termination shock show
that the injection spectrum itself  can deviate significantly from
a simple power law \cite{sla08}, and particle-in-cell simulations
of the acceleration process produce a Maxwellian population with a
power law tail \cite{spi08}.  Any such structure in the injected
particle spectrum imprints itself on the broadband emission of the
entire nebula. The resulting breaks or regions of curvature in the
PWN spectrum, and the frequencies at which they are observed, depend
upon the energy at which features appear in the electron spectrum
as well as the means by which the photons are produced (e.g.
synchrotron radiation or IC emission).  To fully understand the
nature of the particle injection, as well as the long-term evolution
of PWNe, it is thus crucial to study the emission structure over
the entire electromagnetic spectrum.

\begin{figure}[t]
\includegraphics[angle=270,width=11.5cm]{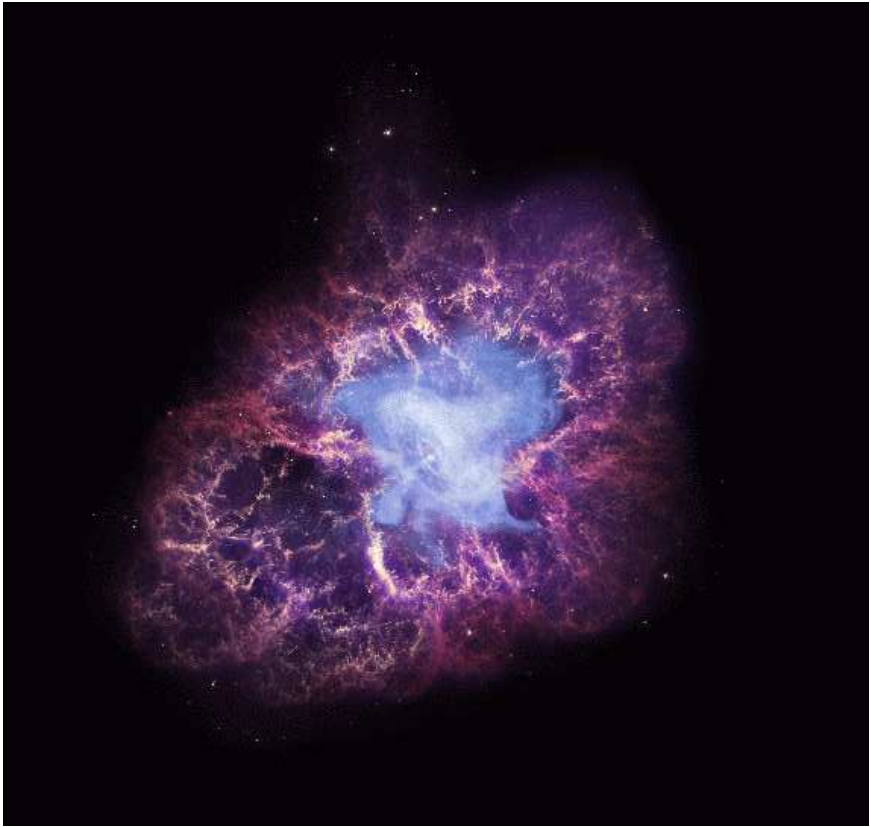}
\caption{
Composite image of Crab Nebula. X-ray emission from \chandra\ is
concentrated in a central jet/torus structure (shown in blue), while
optical emission from {\it HST} (shown in yellow and red), and the
infrared emission from \spitzer\ (shown in purple) dominates the
exterior regions where the nebula sweeps up ejecta material.
}
\label{fig:crab}       
\end{figure}

\section{Dynamical Evolution of PWNe}

The evolution of a PWN within the confines of its host SNR is
determined by both the rate at which energy is injected by the
pulsar and by the density structure of the ejecta material into
which the nebula expands.  The location of the pulsar itself,
relative to the SNR center, depends upon any motion given to the
pulsar in the form of a kick velocity during the explosion, as well
as on the density distribution of the ambient medium into which the
SNR expands. At the earliest times, the SNR blast wave expands
freely at a speed of $\sim (5-10)\times10^3$~\kms, much higher than
typical pulsar velocities of $\sim 400-500$~\kms. As a result, for
young systems the pulsar will always be located near the SNR center.

The energetic pulsar wind is injected into the SNR interior, forming
a high-pressure bubble that expands supersonically into the surrounding
ejecta, forming a shock.  The input luminosity is generally assumed
to have the form (e.g. \cite{ps73})
\begin{equation}
\dot{E} = \dot{E}_0 \left( 1 + \frac{t}{\tau_0}
\right)^{-\frac{(n+1)}{(n-1)}}
\label{eqn_edot_vs_t}
\end{equation}
where
\begin{equation}
\tau_0 \equiv \frac{P_0}{(n-1)\dot{P}_0}
\label{eqn_tau0}
\end{equation}
is the initial spin-down time scale of the pulsar.  Here $\dot{E_0}$
is the initial spin-down power, $P_0$ and $\dot{P}_0$ are the initial spin
period and its time derivative, and $n$ is the so-called ``braking
index'' of the pulsar ($n = 3$ for magnetic dipole spin-down).  The
pulsar has roughly constant energy output until a time $\tau_0$,
beyond which the output declines fairly rapidly with time.
In the spherically symmetric case, the
radius of the PWN evolves as
\begin{equation} R_{PWN} \approx 1.5
\dot{E}_0^{1/5} E_{SN}^{3/10} M_{ej}^{-1/2} t^{6/5} \end{equation}
where $E_{SN}$ is the energy released in the explosion and $M_{ej}$
is the mass of the ejecta \cite{che77}. Thus, at least at early
times when the pulsar input is high, the PWN expansion velocity
increases with time. The sound speed in the relativistic fluid
within the PWN is sufficiently high ($c_s = c/\sqrt{3}$) that any
pressure variations experienced during the expansion are quickly
balanced within the bubble; at early stages, we thus expect the
pulsar to be located at the center of the PWN. The pressure balance
within the PWN results in a termination shock where the energetic
pulsar wind meets the more slowly-expanding PWN. Studies of this 
region of the PWN provide the most direct information available on
particles that are being freshly injected into the nebula.

The geometry of the pulsar system results in an axisymmetric wind
\cite{lyu02}, forming a torus-like structure in the equatorial plane,
along with collimated jets along the rotation axis. The higher
magnetization at low latitudes confines the expansion here to a
higher degree, resulting in an elongated shape along the pulsar
spin axis for the large-scale nebula \cite{beg92,van03}.
This structure is evident in Figure~\ref{fig:crab}, where
X-ray and optical observations of the Crab Nebula clearly reveal
the jet/torus structure surrounded by the elongated wind nebula
bounded by filaments of swept-up ejecta.  A thorough review of the
MHD-driven jet/torus structure is presented by N.~Bucciantini in these
Proceedings.

\begin{figure}[t]
\includegraphics[width=11.5cm]{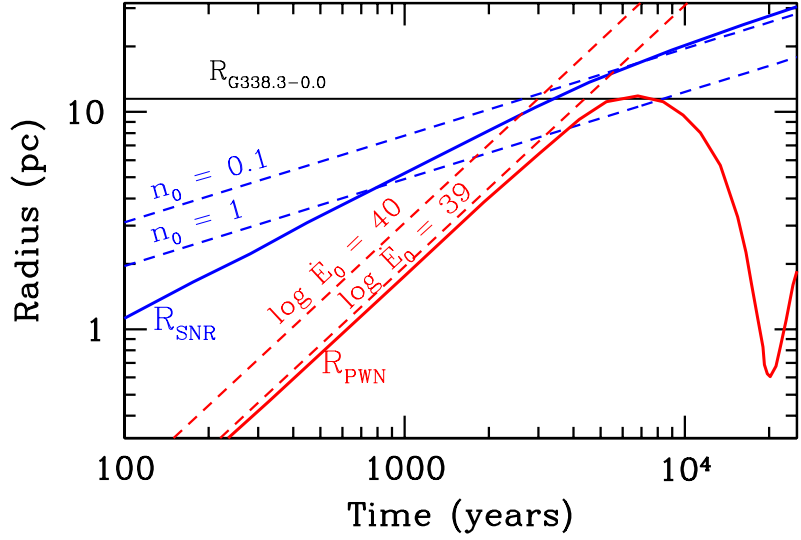}
\caption{
Time evolution of the SNR and PWN radii for a range of values for
the ambient density and initial spin-down power of the pulsar. The
solid curves correspond to models from Gelfand et al. (2009) 
\cite{gsz09} using
$\dot{E_0} = 10^{40} {\rm\ erg\ s}^{-1}$, $M_{ej} = 8 M_\odot$, $n_0
= 0.1 {\rm\ cm}^{-3}$, and $E_{51} = 1$.  See text description for
details. (From \cite{sla10}. Reproduced by permission of the AAS.)
}
\label{fig:dyn_model}       
\end{figure}

Beyond the PWN, the outer blast wave of the SNR drives a shock into the
surrounding interstellar/circumstellar medium (ISM/CSM), forming a
shell of hot gas and compressed magnetic field. As the shell sweeps
up additional mass, and decelerates, a reverse shock (RS) propagates
back into the expanding ejecta. When the SNR has swept up an amount
of mass that is roughly equal to the mass of the ejecta, the evolution
approaches the Sedov phase in which the radius is described
by
\begin{equation}
R_{SNR} \approx 6.2 \times 10^4 \left(\frac{E_{SN}}{n_0}\right)^{1/5} t^{2/5}
\end{equation}
where $n_0$ is the density of the ambient medium. This is illustrated
in Figure~\ref{fig:dyn_model} where dashed lines show $R_{PWN}$ and
$R_{SNR}$ as a function of time using Equations 3 and 4. Here
we have assumed $M_{ej} = 5 M_\odot$ and $E_{SN} = 10^{51}$~erg,
and have used a range of values for $\dot{E}_0$ and $n_0$. As
indicated, eventually the PWN would overtake the SNR boundary under
such conditions.  However, because the initial expansion of the SNR
occurs more rapidly than in the Sedov phase, and the injection rate
from the pulsar is highest at the earliest times, the actual behavior
differs from that shown. In addition, as the reverse shock of the
SNR propagates inward, it eventually reaches the PWN boundary and
begins crushing the nebula. This is indicated in the solid curves
in Figure~\ref{fig:dyn_model}, where the early-phase behavior as
well as the reverse shock propagation is taken into account (see
\cite{gsz09}). Ideally the PWN radius decreases
until the nebula pressure is sufficiently high for the system to
rebound.

The crushing of the PWN results in an increase in the magnetic field.
This causes a rapid burn-off of the most energetic particles in
the nebula. The PWN/RS interface is Rayleigh-Taylor (R-T) unstable, and
is subject to the formation of filamentary structure where the
dense ejecta material is mixed into the relativistic fluid \cite{blo01}.
The nebula subsequently re-forms as the pulsar
injects fresh particles into its surroundings, but a significant
relic nebula of mixed ejecta and relativistic gas will persist.

In cases where the pulsar and its PWN have moved considerably
from the center of the PWN, or in which the SNR has evolved in
a medium of nonuniform density, the reverse shock will interact
with the PWN asymmetrically, encountering one portion of the nebula
well before another. This results in a complex interaction that leaves a
highly distorted relic nebula that may be highly displaced from
the pulsar position \cite{blo01}. 

In the later stages of evolution, the PWN can expand to a very large
size, with a correspondingly lower magnetic field strength. As 
described below, such PWNe may be best identified through their
high-energy $\gamma$-ray emission, with only weak X-ray emission
observed close to the pulsars.

\section{Spectral Evolution of PWNe}

As particles are injected from a pulsar into its PWN, the resulting
emission is determined by the evolved particle spectrum and magnetic
field, as well as the energy density of the ambient photon field.\footnote{In
addition, emission from ejecta gas and dust swept up by the expanding
nebula can be significant -- and even dominant in some wavebands.}
The injected spectrum is often characterized as a power law:
\begin{equation}
Q(E_e,t) = Q_0(t)(E_e/E_0)^{-a}
\end{equation}
The integrated energy in the electron spectrum is then
\begin{equation}
\int Q(E,t) E dE = \frac{1}{(1 + \sigma)} \dot{E}(t)
\end{equation}
where $\sigma$ is the ratio of the spin-down power injected in the form of
Poynting flux to that in the form of particles. It is important to
note one expects the postshock flow of particles to be characterized
by a Maxwellian distribution accompanied by a nonthermal tail -- a
result confirmed by recent particle-in-cell simulations of relativistic
shocks \cite{spi08}. The development of the nonthermal tail depends
on the shock conditions, and the total residence time of the particles
in the acceleration region, which may vary at different locations in the
inner nebula. Thus, it is reasonable to expect that the injected
spectrum may actually deviate considerably from a pure power law form,
a point that recent observations may be beginning to illustrate, 
as we discuss below.

\begin{figure}[t]
\includegraphics[width=11.5cm]{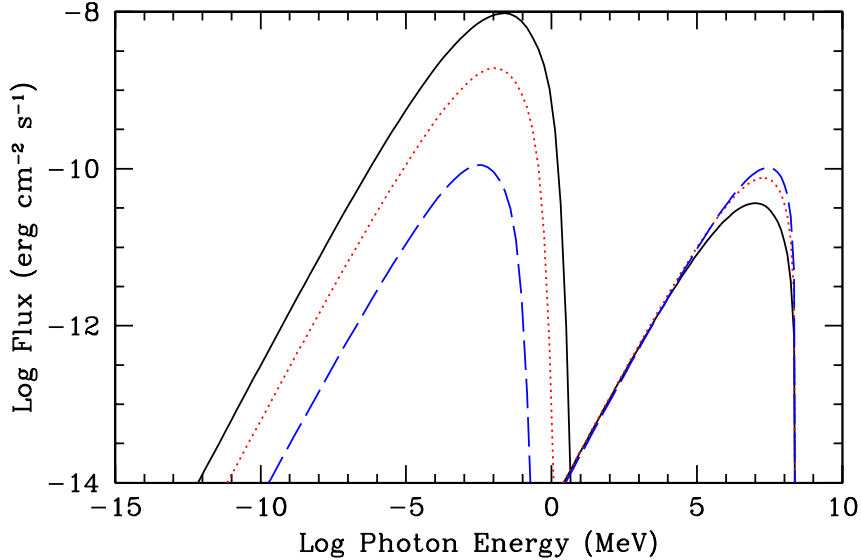}
\caption{
Synchrotron (left) and IC (right) emission (for scattering off of
the CMB) from a PWN at ages of 1000 (solid), 2000 (dotted), and
5000 (dashed) years. Here we have assumed $E_{51} = 1$, $M_{ej} =
8 M_\odot$, and $n_0 = 0.1 {\rm\ cm}^{-3}$ for the SNR evolution,
and $n = 3$, $\dot{E}_0 = 10^{40}{\rm\ erg\ s}^{-1}$, and $\tau_0
= 500$~yr for the pulsar. For the wind, we assume that 99.9\% of
the energy is in the form of electrons/positrons with a power law
spectrum with $\gamma$ = 1.6.
}
\label{fig:pwnrad}       
\end{figure}

The resulting emission spectrum is found by integrating the electron
spectrum over the emissivity function for synchrotron and IC radiation
using, respectively, the nebular magnetic field and spectral density
of the ambient photon field. As illustrated in Figure~\ref{fig:pwnrad},
the build-up of particles in the nebula results in an IC spectrum
that increases with time. The synchrotron flux decreases with time
due to the steadily decreasing magnetic field strength associated
with the adiabatic expansion of the PWN. This behavior is reversed
upon arrival of the SNR reverse shock (not shown in Figure), following
which the nebula is compressed and the magnetic field strength
increases dramatically, inducing an episode of rapid synchrotron
losses. Upon re-expanding, however, IC emission again begins to
increase relative to the synchrotron emission.  At the latest phases
of evolution, when the nebula is very large and the magnetic field
is low, the IC emission can provide the most easily-detected
signature. As described below, such behavior is seen for a number
of PWNe that have been identified based on their emission at TeV
energies, and for which only faint synchrotron emission near the
associated pulsars is seen in the X-ray band.

The broadband spectrum of a PWN, along with the associated dynamical
information provided by measurements of the pulsar spin properties,
and the size of the PWN and its SNR, place very strong constraints
on its evolution and on the spectrum of the particles injected from
the pulsar. Combined with estimates of the swept-up ejecta mass,
this information can be used to probe the properties of the progenitor
star and to predict the long-term fate of the energetic particles
in the nebula. Recent multiwavelength studies of PWNe, combined
with modeling efforts of their evolution and spectra, have provided
unique insights into several of these areas.

\section{Case Studies}

A complete summary of recent results on PWNe is well beyond the scope
of this paper. Below we discuss four distinct systems 
for which recent multiwavelength observations have begun to probe
both the detailed structure and the underlying particle population
in these nebulae.

\subsection{3C 58}

\begin{figure}[t]
\includegraphics[width=11.5cm]{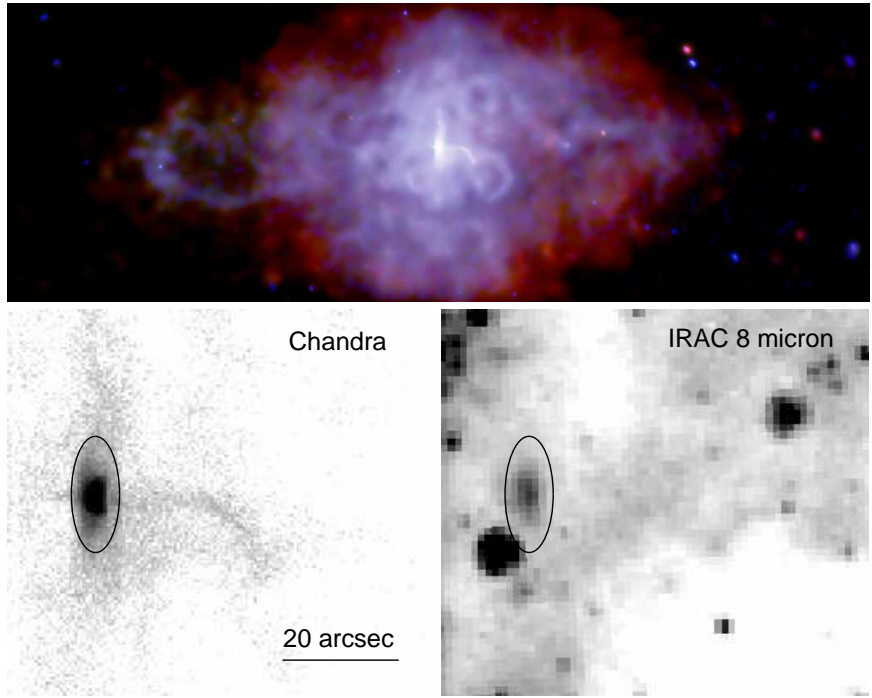}
\caption{
Top: \chandra\ image of 3C 58.  The pulsar is at the center, and
is surrounded by a elongated compact nebula with a curved jet
extending to the west.  A softening of the spectrum with radius is
observed -- an effect resulting from both
synchrotron aging of the electrons and the presence of a soft thermal
shell. Bottom: \chandra\ image of the pulsar in 3C58, and its associated
torus and jet (left) and the IRAC 8~$\mu$m image of the same region.
The elliptical region indicates the torus, and has the same center
and size in each image.
}
\label{fig:3c58_fig}       
\end{figure}

3C~58 is a flat-spectrum radio nebula ($\alpha \approx 0.1$, where
$S_\nu \propto \nu^{-\alpha}$) for which upper limits based on {\em
IRAS} observations indicate a spectral break between the radio and
infrared bands \cite{gs92}. The PWN has often been associated with SN
1181 \cite{sg05}; the low break frequency would then suggest an
extremely large magnetic field ($> 2.5 $mG) if interpreted as the
result of synchrotron break. This has resulted in a number of
different interpretations, including the suggestion that the pulsar
in 3C~58 underwent a rapid decline in its output at some early epoch
\cite{gs92}, the possibility that the low-frequency break is inherent
in the injection spectrum from the pulsar, and the suggestion that
3C~58 is not actually associated with SN~1181, but is an older
nebula \cite{che05,bie06}.

The X-ray emission from 3C~58 (Figure~\ref{fig:3c58_fig}) is
dominated by a power law component, typical of synchrotron emission.
However, a faint thermal component is clearly detected in the outer
regions of the PWN, and also contributes to the interior regions
\cite{boc01,sla04,ghn06}.  The temperature is $\sim 0.25$~keV, and
enhanced abundances of Ne and Mg are observed, indicative of $\sim
0.5 M_\odot$ of ejecta that has been swept up by the PWN.  This is
much larger than the expected mass if the PWN is associated with
SN~1181, and suggests a larger age for the system \cite{che05}.

\begin{figure}[t]
\includegraphics[width=11.5cm]{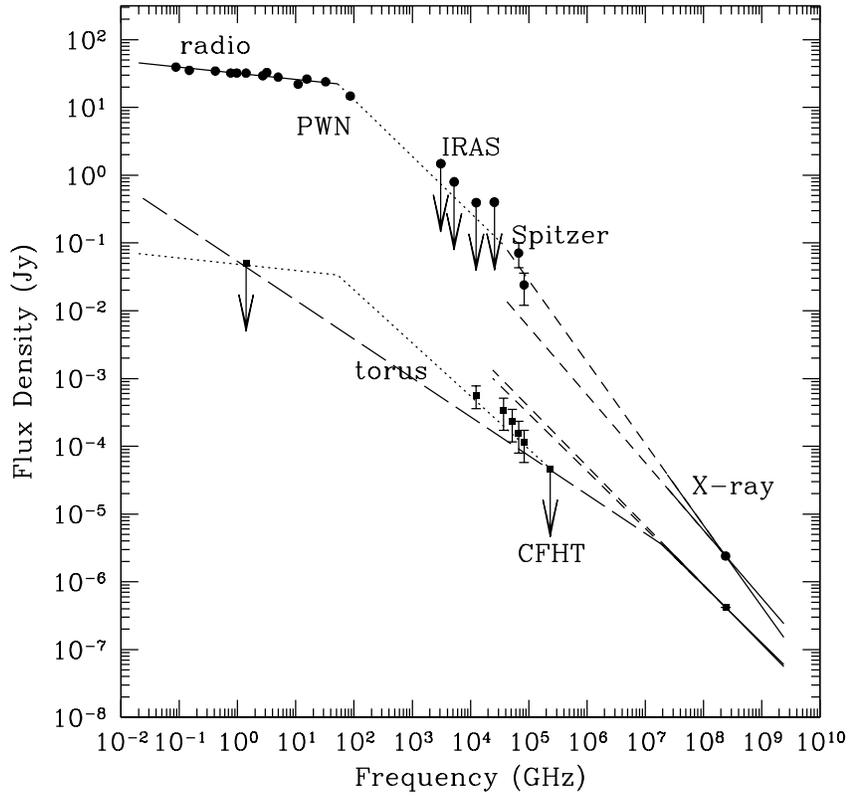}
\caption{
The flux of all of 3C~58 (upper) and its torus (lower), plotted
from the radio to the X-ray band. While the torus is not detected in the
radio band, the IR data require a flattening of the X-ray spectrum when
extrapolated back to the longer wavelength band. (From \cite{sla08}.
Reproduced by permission of the AAS.)
}
\label{fig:3c58_spec}       
\end{figure}

To investigate the evolution of 3C~58, we have carried out \spitzer\
observations using IRAC \cite{sla08}. These observations reveal the
PWN in both the 3.6 and 4.5~$\mu$m bands, representing the first
detection of synchrotron emission from this important young PWN
anywhere in the five decades of frequency separating the radio and
X-ray bands.  The morphology of the IR emission from 3C~58 is
strikingly similar to that seen in the radio and X-ray bands.  The
emission extends all the way to the radio boundaries, indicating
that no synchrotron loss breaks occur below the band, and some
regions of enhanced or diminished emission match well with those
seen in the other bands (notably the large cavity on the eastern
side), suggesting that we are observing primarily synchrotron
radiation.  Optical filaments in 3C~58 \cite{sla04,rf07}, which
presumably originate from supernova ejecta overtaken by the expansion
of the PWN, do not show a good spatial correspondence with the radio
or IR structures, suggesting that the IR emission is not dominated
by dust contributions.  This is similar to results from {\sl Spitzer}
observations of the Crab Nebula \cite{tgw+06}, where emission in
the IRAC band is also identified primarily with synchrotron radiation.

The IRAC data also reveal emission from the torus surrounding the
pulsar in 3C~58 \cite{shm02} in all four bands
(Figure~\ref{fig:3c58_fig}).\footnote{We also show preliminary
results from MIPS observations at 24~$\mu$m where we also detect
the torus.}  Optical emission from the torus is detected as well
\cite{shi08}.  These observations provide new constraints on the
evolution of the particles as they flow from the termination shock
in 3C~58.  There is little question that this emission is synchrotron
in nature; there is insufficient dust in the environment of the
pulsar termination shock to provide a shocked dust component to the
emission. The spectrum (Figure~\ref{fig:3c58_spec}) requires a break
of some sort between the IR and X-ray bands, suggesting that the
synchrotron loss break appears just above the IR band.  Most
importantly, these results indicate that the spectrum of particles
injected into the PWN through the termination shock does not follow
an unbroken power law. As a result, structure in the PWN spectrum
is, at least in part, the imprint of structure from the injection
region.

From Figure~\ref{fig:3c58_spec}, it is clear that additional observations
of 3C~58, and particularly its torus, at longer wavelengths will be
crucially important to understand the nature of the injected particles
and the subsequent long-term evolution of the PWN. Deep observations of
the central regions of other PWNe are clearly of importance as well.

\subsection{Vela X}

\begin{figure}[t]
\includegraphics[width=11.5cm]{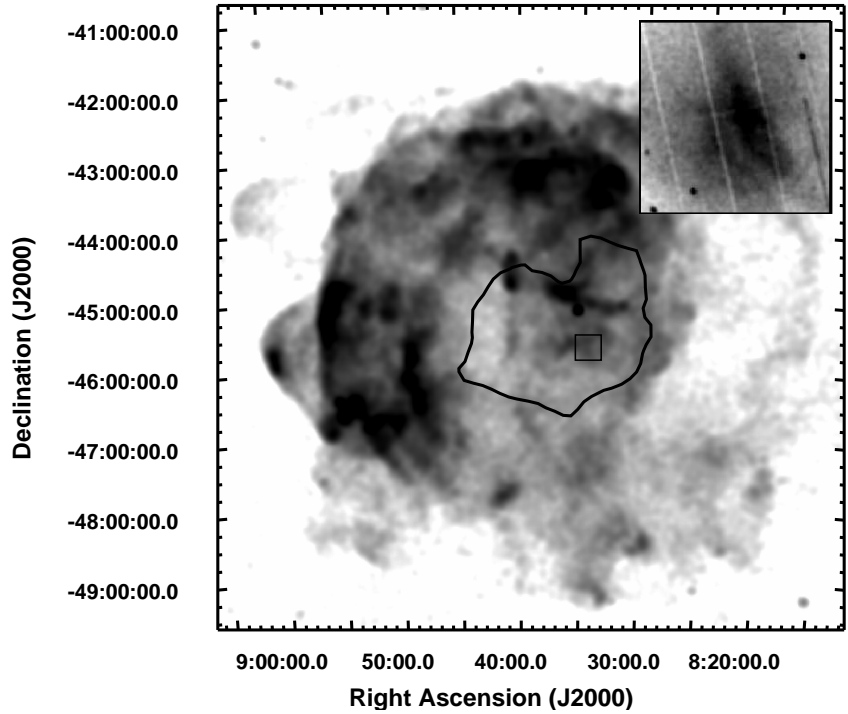} \\
\caption{
\rosat\ PSPC image of Vela SNR. The single contour represents the outer
boundary of the radio nebula Vela X. The point source at the northern extreme
of Vela X is the Vela pulsar and the box inside the PWN indicates the
XMM MOS region of the cocoon shown in the inset.
}
\label{fig:vela_im}       
\end{figure}

Located at a distance of only 280~pc, the Vela SNR houses a young
pulsar that powers the extended nebula Vela X.  This nebula lies
within a limb-brightened shell of thermal X-rays.  The outer shell
is cool, and the exceptionally low foreground absorption ($N_H \sim
1 - 5 \times 10^{20} {\rm\ cm}^{-2}$) allows us to see strong
emission lines from O, Ne, and Mg.  In X-rays, the overall brightness
asymmetry of Vela is evident (Figure~\ref{fig:vela_im}).  The SNR
is much brighter in the northeastern hemisphere, toward the Galactic
plane.  This is apparently the result of large-scale inhomogeneities
in the ISM, with $n_0 \approx 0.06 {\rm\ cm}^{-3}$ in the south
\cite{boc99}, and $n_0 \approx 1 {\rm\ cm}^{-3}$ on the north side
of the SNR \cite{dub98}.

\begin{figure}[t]
\includegraphics[width=11.5cm]{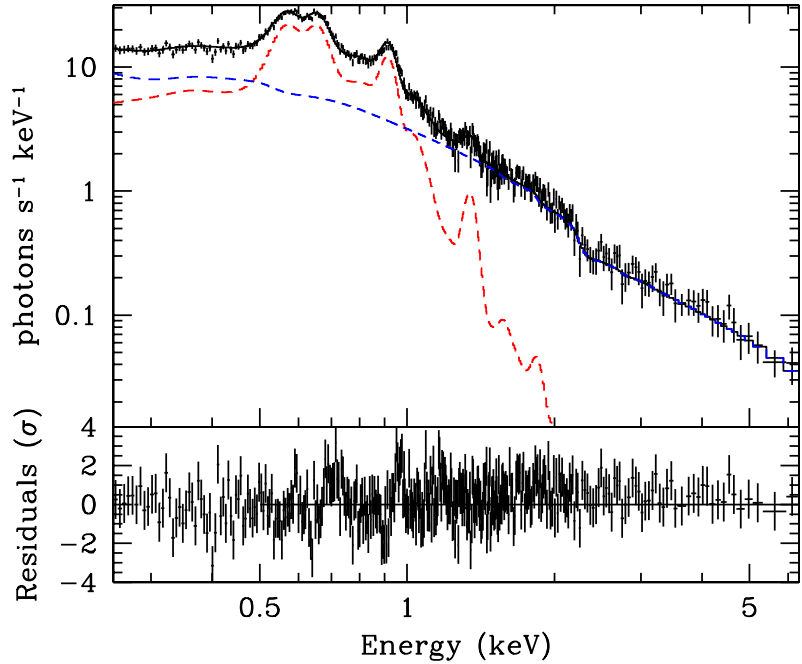} \\
\caption{
{\it XMM-Newton} spectrum from the ``cocoon'' region in Vela X.
The best-fit model, shown in black, is composed of two components --
a thermal plasma with enhanced, ejecta-like abundances (light dashed
curve)
and a power law (dark dashed curve). (From \cite{lsd08}.
Reproduced by permission of the AAS.)
}
\label{fig:velax_xspec}       
\end{figure}

\begin{figure}[t]
\includegraphics[width=11.5cm]{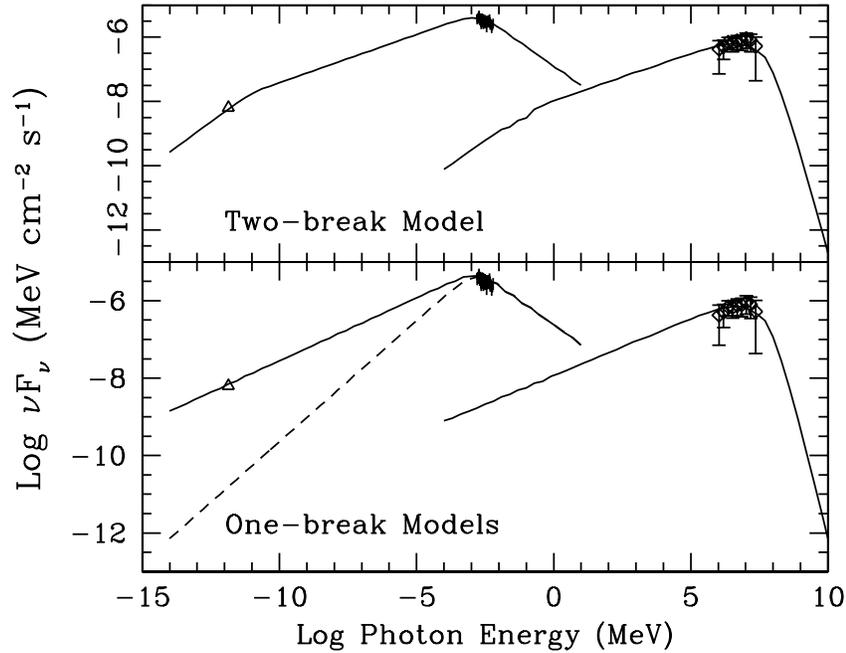}
\caption{
Broadband spectral model consisting of synchrotron emission
in the radio and X-ray bands accompanied by IC emission
in the VHE $\gamma$-ray band. The upper panel shows a model with
two spectral breaks in the electron spectrum. Models with a single break
(lower panel) either underpredict the radio emission, or produce a
radio flux whose spectral index does not agree with observations.
(From \cite{lsd08}. Reproduced by permission of the AAS.)}
\label{fig:velax_model}       
\end{figure}

Radio observations of the PWN \cite{mil80} reveal a morphology
concentrated to the south of the pulsar itself, suggesting that the
nebula has been disrupted by the impact of the reverse shock which
propagated more rapidly from the northeast due to the higher ambient
density in this direction. Higher resolution radio images show a
network of filamentary structure in the PWN \cite{fbm+97}, possibly
formed by R-T instabilities in this interaction with the reverse
shock.  \rosat\ observations of the Vela~X region \cite{mo95} reveal
a large emission structure -- the so-called ``cocoon'' -- extending
to the south of the pulsar.  The region is characterized by a hard
spectrum and appears to lie along a bright elongated radio structure.
{\em ASCA} observations established a two-component X-ray
spectrum with the hard component adequately described by either a
power law or a hot thermal plasma \cite{mo97}.  The PWN is observed
at energies up to $\sim 200$~keV with {\it BeppoSAX} \cite{man05},
and observations with {\it H.E.S.S.} \cite{aha06} reveal extended
VHE $\gamma$-ray emission with the brightest emission concentrated
directly along the cocoon.

Our initial studies of a small region along the cocoon \cite{lsd08}
reveal a bright X-ray structure shown as an inset to
Figure~\ref{fig:vela_im}.  The emission is concentrated into 
several distinct regions, at least some of which appear to be
filamentary structures.  The integrated emission from these regions
is characterized by two distinct components -- a power law with a
spectral index of $\sim 2.2$ and a thermal plasma with enhanced
abundances of O, Ne, and Mg, presumably associated with ejecta that
has been mixed into the PWN upon its interaction with the reverse
shock (Figure~\ref{fig:velax_xspec}).

It is of particular interest that our broadband modeling of the
nonthermal emission from this central region of \velax\ indicates
a disconnect between the radio-emitting particles and those that
produce the X-ray and TeV $\gamma$-ray emission \cite{lsd08}.
A model with a single spectral break between the
bands either underpredicts the radio flux or produces an incorrect
radio spectral index (Figure~\ref{fig:velax_model}), while a model with
two breaks can satisfy the data.  Treating the poorly-characterized
low-energy electron component as a separate population of particles,
de Jager, Slane, \& LaMassa \cite{dsl08} showed that enhanced
IC emission from this component could be expected in
the GeV band.  This has now been confirmed with observations by
{\em AGILE} \cite{pel10} and {\it Fermi} \cite{abdo10}.
Interestingly, a similar excess is observed in {\it Fermi} observations
of HESS~J1640$-$465 \cite{sla10}, another evolved PWN that
appears to have undergone a RS interaction (see Section 4.4). 

The nature of the low-energy particle spectrum in Vela X is poorly
understood, but the {\it Fermi} studies in particular suggest a
difference in the cocoon emission from that of its surroundings;
the LAT emission appears to be concentrated distinctly to the west
of the TeV emission.  Whether or not this emission component is
somehow associated with the reverse shock interaction is not clear.
An \xmm\ Large Project to map a significant portion of Vela X is
underway to investigate the distribution of nonthermal particles
and thermal ejecta in the nebula.

\subsection{G327.1$-$1.1}

\begin{figure}[t]
\includegraphics[width=11.5cm]{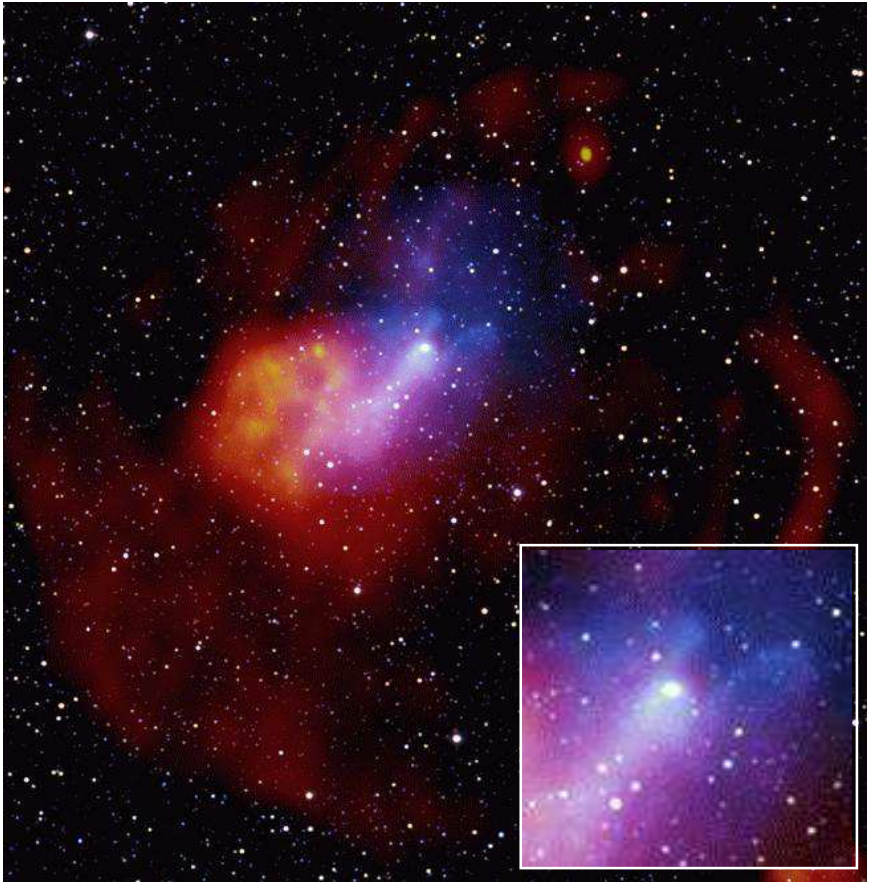}
\caption{
Composite radio (red and orange) and X-ray (blue) image of G327.1$-$1.1.
The outer radio shell defines the SNR boundary, while the bright central
nebula is the PWN. The compact source to the northwest of the nebula
is the neutron star. It is embedded in a cometary-shaped structure
accompanied by a tail of X-ray emission extending into the radio nebula,
as well as prong-like structures (seen in inset) that appear to inflate
a faint bubble in the northwest.
}
\label{fig:g327}       
\end{figure}

G327.1$-$1.1 is a composite SNR with a bright central PWN whose
structure is complex in both the radio and X-ray bands. As shown
in Figure~\ref{fig:g327}, the SNR is characterized by a faint radio
shell accompanied by a bright radio PWN. The PWN is offset from the
SNR center, and a finger-like structure extends toward the
west/northwest. The morphology is suggestive of a PWN that has been
disrupted by an asymmetric reverse shock interaction that has
arrived preferentially from the northwestern direction.

Detailed \chandra\ observations \cite{tem09} appear to confirm this
scenario in detail.\footnote{See also the contribution by T. Temim
in these Proceedings.} A compact X-ray source resides at the tip
of the radio finger, and a trail of nonthermal X-ray emission extends
from the source back into the radio nebula. The compact X-ray source
itself is resolved, with possible evidence of a jet or torus
structure, and the source is embedded in a cometary nebula whose
structure is suggestive of a bow shock.  A pair of prong-like
structures originate from the vicinity of the compact core and
extend out to the west/northwest. Their axes are not aligned with
the compact core, and they do not appear to be jets from the pulsar.
The most unusual feature in G327.1-1.1 is a large bubble-like
structure that extends out from the prongs.  The bubble is very
faint, but it is apparent in images taken with both \chandra\ and
\xmm.

The spectral index in the tail of X-ray emission that extends into
the brighter radio PWN appears to vary from $\sim 1.8$ to $\sim
2.1$, although significantly better statistics are needed to map
this fully.  Faint loop-like structures, possibly associated with
Rayleigh-Taylor filaments or magnetic loops, are evident in the
southern part of the tail and they appear to extend from radio
structures inside the relic PWN. Deeper X-ray observations are
required to better characterize these and, more importantly, to
study the detailed properties of the complex structures seen in
the immediate vicinity of the pulsar, and to investigate the
apparent upstream diffusion of particles that form the bubble-like
structure.

The overall X-ray morphology of the PWN in G327.1-1.1 presents
several challenges. It appears clear from the displacement of the
radio nebula that the PWN has undergone an interaction with the SNR
reverse shock, and that this shock arrived earlier from the northwest,
possibly as a result of pulsar motion in this direction. In such a
scenario, we expect the ongoing pulsar wind production to begin
forming a new PWN around the pulsar, shaped by the surrounding
pressure conditions. Radio emission from the displaced relic nebula
will persist, but the compression from the reverse shock will
temporarily increase the magnetic field in the nebula, causing rapid
synchrotron losses for the more energetic particles. The X-ray
emission is thus expected to be concentrated closer to the pulsar,
as observed. Spectral steeping of the X-ray spectrum in the direction
of the relic nebula is expected if the synchrotron loss timescale
is shorter than the particle flow timescale; mapping this spectral
evolution along the extended X-ray tail thus constrains the conditions
in the relic nebula. In the direction from which the reverse shock
propagated, the structure of the medium is complicated. The density
in the immediate post-shock region is enhanced, but declines
downstream due to adiabatic expansion of the SNR. At least
qualitatively, this could provide the environment in which
freshly-injected wind from the pulsar inflates the observed bubble.

The cometary feature surrounding the putative pulsar complicates
the above picture. The morphology is suggestive of a bow shock that
forms when the pulsar motion exceeds the sound speed in the the
ambient medium, yielding a structure quite different from that of
a static PWN. The wind termination shock in such systems is compressed
in the direction of motion, and extended in the backward direction.
X-ray emission is then observed between the termination shock and
the contact discontinuity, and forms three distinct structures: the
``head'' that surrounds the pulsar; an enhanced region behind the
pulsar, associated with the termination shock; and an elongated
tail where the swept-back wind is concentrated \cite{gae04,buc02}.
For G327.1-1.1, such a geometry would require a significant pulsar
velocity component perpendicular to the plane of the sky, but the
overall velocity that would be required ($770$~\kms) is not
unreasonable.

Given the similarity between G327.1$-$1.1 and Vela, in the context
of both being systems in which an interaction between the PWN and
the SNR reverse shock has occurred, it is of considerable interest
to determine whether or not G327.1$-$1.1, like Vela~X, produces
$\gamma$-rays. Preliminary investigation of \fermi-LAT data
indicates faint emission that is positionally coincident with
G327.1$-$1.1, but further analysis is required to assess this in
detail.

\subsection{HESS J1640$-$465}

\begin{figure}[t]
\includegraphics[width=11.5cm]{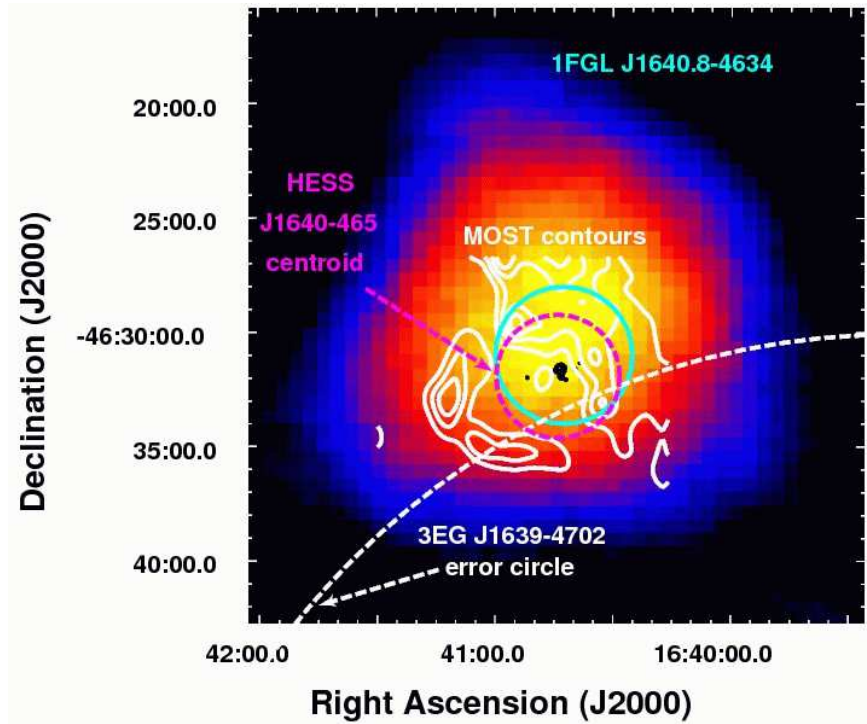}
\caption{
\fermi-LAT image of \pwn1640. The solid (cyan) circle indicates the
uncertainty in the centroid of the \fermi-LAT source, the magenta
dashed circle indicates the 95\% encircled flux distribution of the
H.E.S.S. image, and the white dashed circle indicates that for 3EG
J1639$-$4702. The white contours outline radio emission from
G338.3$-$0.0 while the black contours at the center outline extended
X-ray emission observed with \xmm. A compact X-ray source detected
with \chandra\ resides within the X-ray contours. (From \cite{sla10}.
Reproduced by permission of the AAS.)
}
\label{fig:j1640_im}       
\end{figure}

\pwn1640\ (see Figure~\ref{fig:j1640_im}) is an extended source of
very-high-energy $\gamma$-ray emission discovered with the High
Energy Stereoscopic System (H.E.S.S.) during a survey of the Galactic
plane \cite{aha06b}.  Centered within the radio SNR G338.3$-$0.0
\cite{sg70}, the deconvolved TeV image of the source has an RMS
width of $2.7 \pm 0.5$~arcmin \cite{funk07}.  HI measurements show
absorption against G338.3$-$0.0 out to velocities corresponding to
the tangent point, indicating a distance of at least 8 kpc
\cite{lsg+09}, and thus implying a rather large size for the PWN
($R_{PWN} > 6.4 d_{10} $~pc, where $d_{10}$ is the distance in units
of 10~kpc).  X-ray observations with \xmm\ \cite{funk07} and \chandra\
\cite{lsg+09} establish the presence of an accompanying X-ray nebula
and an X-ray point source that appears to be the associated neutron
star. The point source is offset from the center of the PWN, and
both are offset from the center of the SNR, suggesting that an
asymmetric interaction with the SNR reverse shock has occurred.

\begin{figure}[t]
\includegraphics[width=11.5cm]{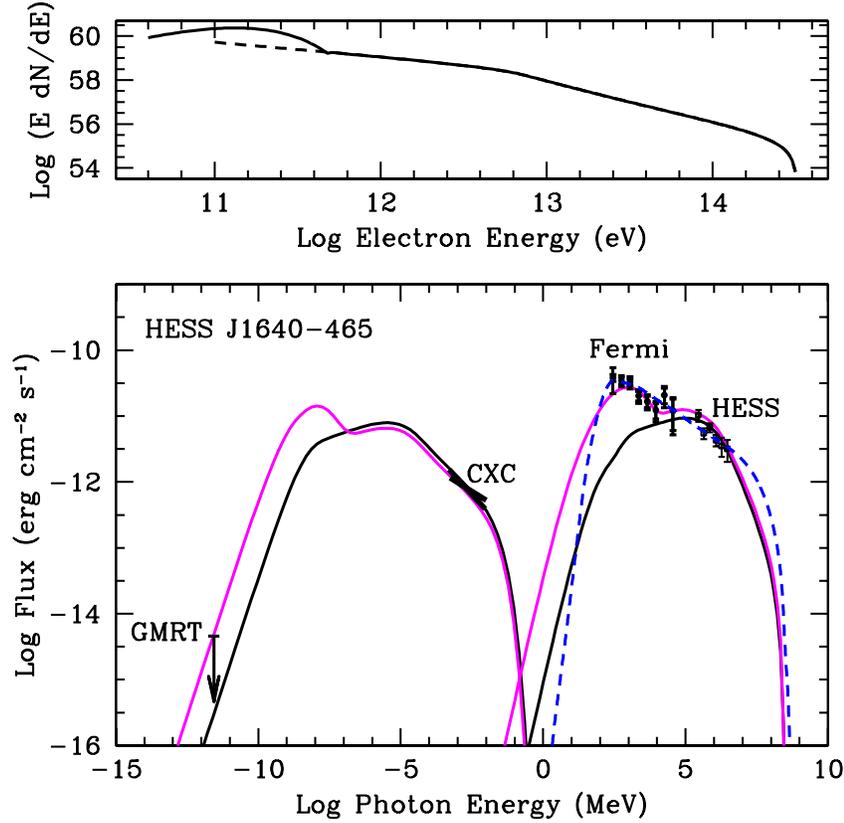}
\caption{
Electron spectrum (upper) and broadband emission model (lower) for
\pwn1640\ assuming the evolutionary history described in the text.
The black curves represent a PWN with an age $T = 10$~kyr, and $B(T)
= 5 \mu$G, assuming $\dot{E_0} = 4 \times 10^{36} {\rm\ erg\ s}^{-1}$
and an injection spectrum with $\sigma = 10^{-3}$, $\gamma = 2.5$,
and $E_{\rm min} = 115$~GeV. 
The light curves represent the
scenario with a low-energy Maxwellian electron component replacing
the low-energy portion of the electron power-law spectrum. The mean
temperature for the IR and optical photon fields are 15~K and 5000~K,
respectively, and the energy densities relative to the CMB are 4
and 1.15. The dashed curve in the upper panel represents the truncated
portion of the power law that was replaced by a Maxwellian. The
dashed curve in the lower panel represents a model for which
all of the $\gamma$-ray emission results from pion decay. (From
\cite{sla10}. Reproduced by permission of the AAS.)
}
\label{fig:j1640_mod}       
\end{figure}

We have investigated \fermi-LAT data acquired from the region
surrounding \pwn1640\ and detect the source with high significance
\cite{sla10}.  The spectrum is well-described by a power law with
$\Gamma = 2.30 \pm 0.09$ and a $F(>100{\rm\ MeV}) = 2.8 \times
10^{-7}{\rm\ photons\ cm^{-2}}{\rm\ s}^{-1}$.  We have modeled the
emission assuming a 1-zone model in which particles are injected
into the nebula with a simple power law distribution.  We use a
radius of $R_{SNR} \sim 11.6 d_{10}$~pc for \snr338, based on radio
observations.  The observed extent of \pwn1640\ constrains the
radius of the PWN to $R_{PWN} > 6.4 d_{10}$~pc. As indicated in
Figure~\ref{fig:dyn_model}, where a horizontal line indicates the
radius of \snr338, reasonable values for the SNR and PWN parameters
indicate that the SNR reverse shock has almost certainly begun to
impact the PWN.

The broadband emission model results are shown in
Figure~\ref{fig:j1640_mod} where we plot the \fermi\ and H.E.S.S
spectra along with the radio upper limit from GMRT observations
\cite{gcd+08} and spectral confidence bands derived from
\chandra\ \cite{lsg+09}.  The black curves represent the model
prediction for the synchrotron (left) and IC (right) emission that
best describes the X-ray and TeV $\gamma$-ray spectra, similar to
results from \cite{lsg+09}; the parameters for the model
are summarized in the caption.  As seen in Figure~\ref{fig:j1640_mod},
this model significantly underpredicts the observed \fermi-LAT
emission.  Our spectral fits can formally accommodate up to about
$\sim 20\%$ of the observed flux in a pulsar-like component
characterized by a power law with an exponential cutoff energy
between 1 and 8 GeV, and there are several known radio pulsars
located within the error circle of \pwn1640 that could potentially
produce observable $\gamma$-ray emission. Even in this case, however,
the \fermi-LAT emission still greatly exceeds the predicted flux from
\pwn1640.

As described in Section~4.2, simple power-law models for the particles
in Vela X, another evolved PWN, fail to reproduce the observed
broadband spectrum. The presence of an excess population of low-energy
electrons is inferred, and models for the IC scattering of photons
from this population predict an excess of $\gamma$-rays in the GeV
range.  Motivated by these results from Vela~X, we modified the
evolved power law spectrum from our model for \pwn1640\ by truncating
the lower end of the power law and adding a distinct low-energy
component.  Based on results from simulations of shock acceleration
\cite{spi08}, we chose a Maxwellian distribution for this population.
Our resulting (ad hoc) particle spectrum is shown in the upper panel
in Figure~\ref{fig:j1640_mod}, and the resulting broadband emission
is shown in the lighter (magenta) curves in the lower panel. Here
we have adjusted the normalization of the Maxwellian to reproduce
the emission in the \fermi-LAT band, which is produced primarily
by upscattered infrared (IR) photons from local dust.  We find a
mean value of $\gamma \approx 2 \times 10^5$ for the electrons in
the Maxwellian component, and roughly 5\% of the total electron
energy in the power law tail, consistent with results from
particle-in-cell simulations.  Recent work by Fang \& Zhang (2010)
\cite{fz10} uses a similar input distribution to successfully model
the emission for several PWNe including \pwn1640.  However, their
results for \pwn1640\ underpredict the observed GeV emission from
this source, apparently due to use of a slightly lower bulk Lorentz
factor and a larger fraction of the total energy in the power law
tail than we have used in this analysis.

An alternative scenario for the $\gamma$-ray emission is that it
arises from the SNR itself, and not the PWN.  The dashed blue curve
in Figure~\ref{fig:j1640_mod}  represents a model for the emission
from the collision of protons accelerated in the SNR with ambient
material, leading to $\gamma$-rays from the production and subsequent
decay of neutral pions. Assuming a shock compression ratio of 4 and
that 25\% of the total supernova energy appears in the form of
relativistic protons, an ambient density $n_0 \approx 100 {\rm\
cm}^{-3}$ is required to produce the model shown in the Figure. This
is much higher than can be accommodated for the observed size of
the SNR and the lack of observed thermal X-ray emission from the
SNR.  Such high densities are found in dense molecular clouds,
suggesting that the $\gamma$-rays could be produced by particles that
stream away to interact with high-density material outside the SNR.
However, only the most energetic particles can escape the acceleration
region, which is in conflict with the proton spectrum we require
to match the data. Moreover, the observed TeV emission appears to
originate from within the SNR boundaries, making such an escaping-particle
scenario appear problematic.  Based on this, along with the lack
of a spectral cutoff that might suggest emission from a central
pulsar, we conclude that the GeV $\gamma$-ray emission most likely
arises from the PWN.

\section{Summary}
The broadband spectra of PWNe provide information about both the
structure and evolution of these objects. New multiwavelength observations
have begun to probe PWNe from the sites of particle injection to
the ejecta-laden outer boundaries, providing crucial input for modeling
these systems. Observations in the $\gamma$-ray band have uncovered
previously unknown systems in the late phase of evolution, while X-ray
observations continue to provide detailed information about the geometry
and the composition of the pulsar winds. These observations 
continue to inform theoretical models of relativistic shocks which, in
turn, have broad importance across the realm of high-energy astrophysics.
At the same time, these recent results have pointed the way to new
and deeper observations of PWNe across the electromagnetic spectrum.

\begin{acknowledgement} 

I would like to acknowledge the considerable contributions of many
colleagues to the work described here, including Bryan Gaensler,
David Helfand, Stephen Reynolds, Okkie de Jager, and Stefan Funk.
In particular, I would like to thank Yosi Gelfand, Tea Temim, Daniel
Castro, Stephanie LaMassa, and Anne Lemiere who, as students as postdocs,
led much of the work described here.

This work was supported in part by NASA contract NAS8-03060, NASA grants
NNX09AT68G, and NNX09AP99G, and Spitzer RSA 1375009.

\end{acknowledgement}
%

%
%

\begin{thebibliography}{99.}%

\bibitem{abdo10} 
 Abdo, A.~A., et al.\ 2010, \apj, 713, 146

\bibitem{aha06}
 Aharonian, F. A. et al.\ 2006, A\&A, 448, L43

\bibitem{aha06b}
Aharonian, F.~A. et al.\ 2006, A\&A, 636, 777

\bibitem{beg92}
 Begelman, M.~C., \& Li, Z.-Y.\ 1992, \apj, 397, 187 

\bibitem{bie06}
 Bietenholz, M. ~F., Kassim, N. E., \& Weiler, K. W. 2001, \apj, 560, 772

\bibitem{blo01}
 Blondin, J.~M., Chevalier, R.~A., \& Frierson, D.~M.\ 2001, \apj,
 563, 806 

\bibitem{boc99}
 Bocchino et al. 1999, \aap, 342, 839

\bibitem{boc01}
 Bocchino, F. et al.\  2001, A\&A, 369, 1078

\bibitem{buc02} 
 Bucciantini, N.\ 2002, \aap, 387, 1066

\bibitem{che77}
 Chevalier, R. A. 1977, in Astrophysics and Space Science Library,
 Vol. 66, Supernovae, ed. D. N.  Schramm, 53

\bibitem{che05}
 Chevalier, R.~A.\ 2005, \apj, 619, 839

\bibitem{dsl08}
 de Jager, O. C., Slane, P. O., \& LaMassa, S. M.\ 2008, \apj, 689, 125

\bibitem{dub98}
 Dubner et al. 1998, AJ, 116, 813

\bibitem{fz10}
 Fang, J. \& Zhang, L. 2010, A\&A, 515, A20

\bibitem{fbm+97}
 Frail, D. A. et al.\ 1997, \apj, 475, 224

\bibitem{funk07}
 Funk, S., et al. 2007, ApJ, 267, 517

\bibitem{gae04}
 Gaensler, B.~M. et al. \ 2004, \apj, 616, 383

\bibitem{gae06}
 Gaensler, B.~M., \& Slane, P.~O.\ 2006, ARA\&A, 44, 17 

\bibitem{gsz09}
 Gelfand, J.~D., Slane, P.~O., \& Zhang, W.\ 2009, \apj, 703, 2051 

\bibitem{gcd+08}
 Giacani, E. et al. 2008, AIPC, 1085, 234

\bibitem{gs92}
 Green, D. A. \& Scheuer, P. A. G.\ 1992, \mnras, 258, 833

\bibitem{ghn06}
 Gotthelf, E.~V., Helfand, D.~J., \& Newburgh, L.\ 2006, \apj, 654, 267

\bibitem{lsd08}
 LaMassa, S. M., Slane, P. O., \& de Jager, O. C.\ 2008, \apj, 689, 121L

\bibitem{lsg+09}
 Lemiere, A., Slane, P., Gaensler, B.~M., \& Murray, S. 2009, ApJ, 706, 1269

\bibitem{lyu02}
 Lyubarsky, Y.~E.\ 2002, \mnras, 329, L34 

\bibitem{man05}
 Mangano, V. et al. 2005, A\&A, 436, 917

\bibitem{mo95}
 Markwardt \& \"{O}gelman 1995, Nature, 375, 40

\bibitem{mil80}
 Milne, D. K.\ 1980, \aap, 81, 293

\bibitem{mo97}
 Markwardt \& \"{O}gelman 1997, \apj, 480, L13

\bibitem{ps73}
 Pacini, F. \& Salvati, M.\ 1973, \apj, 186, 249

\bibitem{pel10}
 Pellizzoni, A., et al. 2010, Sci, 327, 663

\bibitem{ren84}
 Reynolds, S.~P., \& Chevalier, R.~A.\ 1984, \apj, 278, 630 

\bibitem{rf07}
 Rudie, G. C., \& Fesen, R. A. 2007, RMxAC, 30, 90

\bibitem{sg70}
 Shaver, P. A. \& Goss, W. M. 1970, AuJPA, 14, 133

\bibitem{shi08}
 Shibanov, Y. A.\ 2008, \aap, 486 273

\bibitem{shm02}
 Slane, P., Helfand, D.~J., \& Murray, S.~S. 2002, ApJ, 571, L45

\bibitem{sla04}
 Slane, P., Helfand, D.~J., van der Swaluw, E., \& Murray, S.~S.\ 
 2004, \apj, 616, 403 

\bibitem{sla08}
 Slane, P., et al. 2008, \apj, 676, L33

\bibitem{sla10}
 Slane, P., et al. 2010, \apj, 720, 266

\bibitem{spi08}
 Spitkovsky, A.\ 2008, \apj, 682, L5

\bibitem{sg05}
 Stephenson, F.~R., \& Green, D.~A.\ 2005, Historical Supernovae and Their
 Remnants (Oxford University Press, USA)

\bibitem{tgw+06}
 Temim, T. et~al.\ 2006, \aj, 132, 1610

\bibitem{tem09} 
 Temim, T., et al.\ 2009, \apj, 691, 895 

\bibitem{van03}
 van der Swaluw, E.\ 2003, \aap, 404, 939 

%
\end{thebibliography}
%

\end{document}